\documentclass[sigconf]{acmart}
\AtBeginDocument{%
  \providecommand\BibTeX{{%
    \normalfont B\kern-0.5em{\scshape i\kern-0.25em b}\kern-0.8em\TeX}}}

\copyrightyear{2024}
\acmYear{2024}
\setcopyright{acmlicensed}
\acmConference[WWW '24 Companion]{Companion Proceedings of the ACM Web Conference 2024}{May 13--17, 2024}{Singapore, Singapore}
\acmBooktitle{Companion Proceedings of the ACM Web Conference 2024 (WWW '24 Companion), May 13--17, 2024, Singapore, Singapore}
\acmDOI{10.1145/3589335.3651499}
\acmISBN{979-8-4007-0172-6/24/05}

\usepackage{multirow}
\usepackage{adjustbox}
\usepackage{nicematrix}
\usepackage{makecell}

\begin{document}

\title{End-to-End Graph-Sequential Representation Learning for Accurate Recommendations}

\author{Vladimir Baikalov}
\email{Vladimir.Baikalov@skoltech.ru}
\orcid{0009-0009-4864-2305}
\affiliation{%
  \institution{Skolkovo Institute of Science and Technology}
  \streetaddress{Nobel St. 3, Skolkovo Innovation Center}
  \city{Moscow}
  \country{Russia}
  \postcode{143025}
}

\author{Evgeny Frolov}
\email{frolov@airi.net}
\orcid{0000-0003-3679-5311}
\affiliation{%
  \institution{Artificial Intelligence Research Institute (AIRI)}
  \city{Moscow}
  \country{Russia}
}
\affiliation{%
  \institution{Skolkovo Institute of Science and Technology}
  \streetaddress{Nobel St. 3, Skolkovo Innovation Center}
  \city{Moscow}
  \country{Russia}
  \postcode{143025}
}

% \renewcommand{\shortauthors}{Baikalov and Frolov}

\begin{abstract} 
Recent recommender system advancements have focused on developing sequence-based and graph-based approaches. Both approaches proved useful in modeling intricate relationships within behavioral data, leading to promising outcomes in personalized ranking and next-item recommendation tasks while maintaining good scalability. However, they capture very different signals from data. While the former approach represents users directly through ordered interactions with recent items, the latter aims to capture indirect dependencies across the interactions graph. This paper presents a novel multi-representational learning framework exploiting these two paradigms' synergies. Our empirical evaluation on several datasets demonstrates that mutual training of sequential and graph components with the proposed framework significantly improves recommendations performance.

\end{abstract}


\begin{CCSXML}
<ccs2012>
<concept>
<concept_id>10002951.10003317.10003347.10003350</concept_id>
<concept_desc>Information systems~Recommender systems</concept_desc>
<concept_significance>500</concept_significance>
</concept>
 </ccs2012>
\end{CCSXML}

\ccsdesc[500]{Information systems~Recommender systems}
\keywords{Recommender Systems, Sequential Learning, Graph Neural Networks, Contrastive Learning}

\begin{teaserfigure}
  \includegraphics[width=\textwidth]{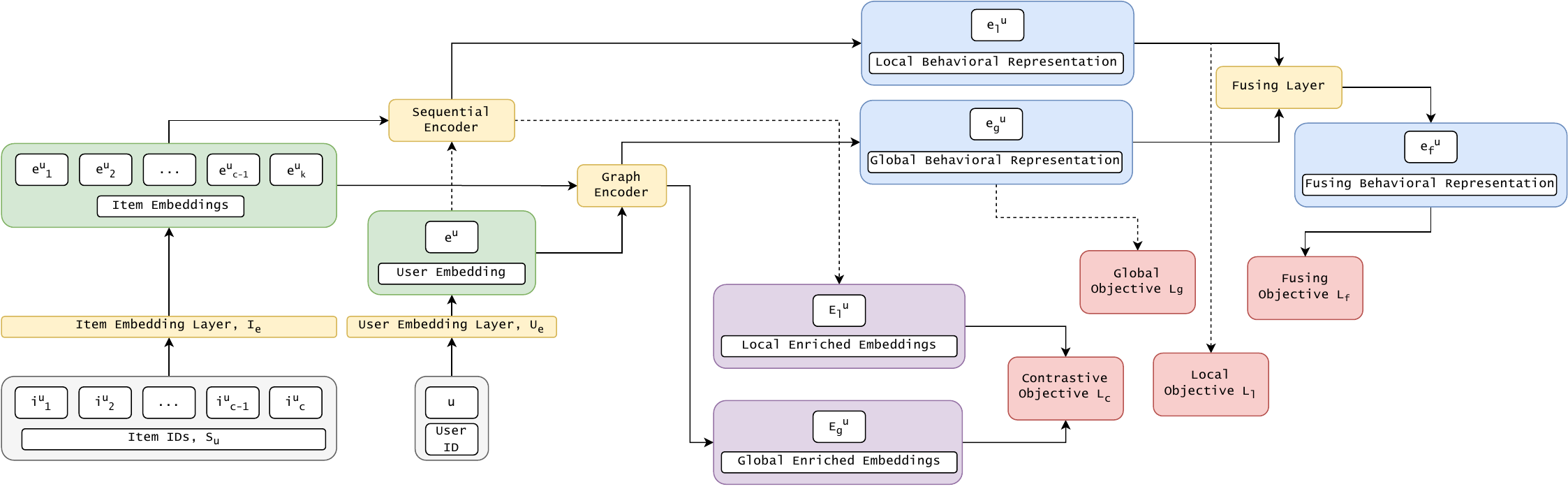}
  \caption{The proposed MRGSRec model consists of nine blocks, colored in yellow and red: (i) Item Embedding Layer and (ii) User Embedding Layer that encodes input item and user IDs. (iii) Sequential Encoder and (iv) Graph Encoder that aggregate information leveraging different data representations. (v) Fusing Layer capable of mixing representations from both encoders. (vi) Local Objective, (vii) Global Objective, (viii) Fusing Objective, and (ix) Contrastive Objective used for model training}
  \Description{MRGSRec model overview}
  \label{fig:model_overview}
\end{teaserfigure}


\maketitle

\section{Introduction}

The primary objective of recommender systems is to offer users personalized recommendations aligned with their interests by leveraging past engagements between users and items. In recent years, deep learning-based solutions have emerged as dominant players in the evolution of recommender systems, causing the rapid development of diverse techniques and methods~\cite{survey_1, survey_2}. 
These solutions can construct accurate and expressive representations of users and items and address various recommendations tasks~\cite{sasrec, bert4rec, light_gcn}. In most cases, input data for such models may be represented as a user-specific sequence of recently interacted items. Leveraging only this information, however, has its limitations: concerned with individual user interactions, this approach captures patterns only for direct dependencies, which may lack sensitivity to a broader context provided by additional interdependencies across entities.
On the contrary, Graph Neural Networks~(GNNs)~\cite{ngcf, light_gcn} provide a powerful way of explicitly modeling information through indirect dependencies, but adapting them to sequential setup can be cumbersome.
Each approach therefore addresses the same problem of learning user preferences by leveraging patterns from independent data representations. To our knowledge, only one study has attempted to integrate both representations within a single model~\cite{mclsr}. However, it was constrained by a sequential encoder architecture. In contrast, we aim to design a general approach, allowing the useage of various versions of sequential and graph-based components, constrained only by encoder signatures. Our contributions are as follows:

\begin{itemize}
    \item We propose \textbf{MRGSRec} (\textbf{M}ulti-\textbf{R}epresentation \textbf{G}raph-based and \textbf{S}equence-based approach for \textbf{Rec}ommendations), novel architecture design that integrates graph-based and sequential representations into an end-to-end framework;
    \item We present a scheme for knowledge exchange between different representations based on contrastive learning;
    \item We assemble a composite loss function that maximizes the cumulative effect of all components of the architecture.
\end{itemize}

\section{Related work}
We categorize existing models into three types: sequence-based, graph-based, and multi-representational (models that integrate both categories into one algorithm).

\paragraph{Sequential Recommendations}

Transformer architectures~\cite{attention} have proven to be effective across many domains~\cite{gpt, vit, detr}, including their utility in sequence modeling. Their unique properties have contributed to their success in various applications, including recommender systems, leading to the development of top-performing methods.
% Given input data in the form of sequences, the transformers~\cite{attention} emerge as one of the best architectures for recommender systems problems, demonstrating its efficacy across various domains~\cite{gpt, vit, detr}. This architecture also has garnered attention for its adeptness in capturing sequential signals from input data.
Examples include SASRec~\cite{sasrec} and BERT4Rec~\cite{bert4rec} demonstrating state-of-the-art performance. The effectiveness of these attention-based architectures in modeling sequential data sparked many further improvements and adaptations aimed at enhancing recommendations quality~\cite{cl4srec, duorec, sasrec_ce}. Notably, the usage of cross-entropy loss for SASRec architecture (SASRec+)~\cite{sasrec_ce} has demonstrated superior performance compared to other techniques. 

\paragraph{Graph-based Recommendations}

An alternative approach to match user preferences involves using graph representations from user-item interactions. Graph Convolution Networks~(GCNs), notably NGCF~\cite{ngcf} and LightGCN~\cite{light_gcn}, have been recognized as highly effective solutions within this domain, with NGCF being foundational and LightGCN serving as both an ancestor and an enhancement.

\paragraph{Multi-Representational Recommendations}

Both previous ideas have their limitations: sequential recommendations excel at extracting patterns from user-specific interactions but overlook broader interactions. Conversely, GCNs grasp the indirect interdependencies but struggle to leverage them within user-specific history as precise as sequential models. To the best of our knowledge, only one approach, namely MCLSR~\cite{mclsr} attempts to bridge this gap by using both paradigms, however, the fusing between different representations is used only during the training process by propagating patterns from indirect interdependencies into item representations, while during the inference only sequential part is leveraged. Also, MCLSR uses an architecture with limited capacity for the sequential part of the model. MRGSRec enhances the idea from MCLSR by utilizing both sequence-based and graph-based representations during the whole training and inference process and providing flexible choices of different encoders for both sequence-based and graph-based parts by having only signature constraints. By combining various architectures, this design offers an end-to-end solution.

\section{Proposed approach}

This section presents the problem statement for the considered recommendation scenario and explains in detail the components of the MRGSRec framework, providing insights into their roles in enhancing model quality. Model overview is shown in Fig.~\ref{fig:model_overview}.

We consider a standard next-item prediction task. The goal is to establish relations between a set $\mathcal{U}$ of $M$ users and a set $\mathcal{I}$ of $N$ items in a sequential manner, i.e., for a given user $u \in \mathcal{U}$ described by a sequence of the previously interacted items \mbox{$S_u = (i^u_1,\, i^u_2,\, \dots,\, i^u_{|S_u|})$}, $i^u_k \in \mathcal{I}$, the task is to predict the next item $i^u_{|S_u| + 1}$ the user is likely to interact with. Below we outline the components of our solution.

\subsection{Embedding layers}

Given a user ID $u$ and an interacted sequence $S_u$ of item IDs, we keep only the last $c$ interactions from $S_u$. We then convert the user and remaining item IDs into embeddings using learnable embedding layers $U_{e}: \mathbb{N} \rightarrow \mathbb{R}^d$ and $I_{e}: \mathbb{N} \rightarrow \mathbb{R}^d$ for user ID and item IDs respectively. This process produces a user embedding $e^u = U_{e}(u) \in \mathbb{R}^d$ and a sequence of item embeddings $\left(e^u_1, e^u_2, \dots, e^u_{c}\right) = E^u \in \mathbb{R}^{c \times d}$, where $e^u_k = I_{e}(i^u_k)$ and $d$ denotes the embedding size. To enhance item representation with position-specific information, we add learnable positional embeddings similar to existing approaches~\cite{sasrec, bert4rec, cl4srec, duorec}.

\subsection{Sequential Encoder}

To capture the direct interactions, we utilize a sequential encoder block within the MRGSRec framework. This block is tasked with producing two representations: an aggregated \emph{local behavioral representation} and \emph{local enriched embeddings} for all input items, with dimensions $\mathbb{R}^d$ and $\mathbb{R}^{c \times d}$ respectively. Without integrating user-specific data, the encoder output lacks personalized context, producing user-agnostic output. We address this limitation by concatenating $e^u$ and $E^u$ before applying the encoder, allowing the sequential encoder to enrich each item representation with user-specific context. As specified in the  equation above, then we extract \emph{local behavioral representation} $e_l^{u}$ and \emph{local enriched embeddings} $E_l^{u}$:
\begin{equation}
    e_l^{u} \, \| \, E_l^{u} = \text{SequentialEncoder}\left(
    e^u \, \| \, E^u \right),
    \label{eq:transformer}
\end{equation}
where SequentialEncoder is an encoder part from the transformer architecture~\cite{attention}, that processes an input matrix of size $(c + 1) \times d$ to produce an enriched representation of the same size.

\subsection{Graph Encoder}

To incorporate the ability to capture indirect interdependencies, we propose employing GCNs, a method used in various prior works~\cite{survey_2, ngcf, light_gcn}. Within the MRGSRec framework, applying a graph encoder requires similar constraints as discussed in the sequential encoder section. Using GCNs involves introducing two input matrices: the initial representation matrix for all users and items $E^{(0)}_g \in \mathbb{R}^{(M + N) \times d}$ and the graph adjacency matrix $A \in \mathbb{R}^{(M + N) \times (M + N)}$. To construct $E^{(0)}_g$, we utilize embeddings for all users and items from the previously defined embedding layers $U_{e}$ and $I_{e}$ by concatenating them. For adjacency matrix initialization, we follow the approach from prior works~\cite{ngcf, light_gcn}. Initially, we introduce the interaction matrix $R \in \mathbb{R}^{M \times N}$, in which elements $r_{ij}$ are set to 1 when there is an interaction between user $i$ and item $j$, and 0 otherwise. Using matrix $R$, we can define matrix $A$ in the block notation:
\begin{equation}
    A = \begin{pmatrix}
        0 & R \\
        R^{\top} & 0 \\
    \end{pmatrix}.
\end{equation}
To apply graph encoder, we use $E^{(0)}_g$ and $A$ in the recursive equation:
\begin{equation}
    E^{(k)}_g = \text{GraphEncoder}\left(E_g,\, A \right) = D^{-\frac{1}{2}} \, A \, D^{-\frac{1}{2}} \, E^{(k - 1)}_g \, ,
    \label{eq:graph_encoder}
\end{equation}
where $A$ denotes the graph's adjacency matrix, $D_{ii} = \sum_{j = 1}^{M + N}{a_{ij}}$ stands for the diagonal degree matrix of $A$, $k$ represents the number of graph encoder layers, and $E^{(k)}_g$ signifies the output of the graph encoder, which provides enriched representations for all users and items. To obtain the \emph{global behavioral representation} $e^u_g \in \mathbb{R}^{d \times 1}$ and \emph{global enriched embeddings} $E^u_g \in \mathbb{R}^{c \times d}$ following the graph encoder's execution, we derive the final representation from $E^{(k)}_g$ using the IDs of given user $u$ and each item in $S_u$.

\subsection{Fusion Layer}

After processing input data with both sequential and graph encoders, we obtain pairs of embeddings $e^u_l$ and $e^u_g$ for local and global behavioral representations, respectively. To integrate information from both encoders into our final representation, we utilize a shallow feed-forward block as described in Eq.~\ref{eq:fusing}:
\begin{equation}
    e^u_f = W_2 \, \text{ReLU} \left( W_1  ( e^u_l \| e^u_g ) \right) ,
    \label{eq:fusing}
\end{equation}
where $e^u_f \in \mathbb{R}^{d}$ denotes the fused behavioral representation. $W_1 \in \mathbb{R}^{4d \times 2d}$ and $W_2 \in \mathbb{R}^{d \times 4d}$ are learnable projection matrices, and $\|: \mathbb{R}^{d},\, \mathbb{R}^{d} \rightarrow \mathbb{R}^{2d}$ represents the concatenation function.

\subsection{Loss Functions}

The essential component of MRGSRec, enabling end-to-end usage of both contexts, lies in its loss functions. Inspired by the cross-domain recommender systems~\cite{cross_domain}, we construct objectives for every intermediate step within our framework. For context-specific behavior representations, we use local and global objectives $\mathcal{L}_{l}$ and $\mathcal{L}_{g}$ which are the same as the downstream loss functions from the chosen encoder architecture (e.g., next item prediction from SASRec~\cite{sasrec} for sequential encoder and BPR-loss plus regularization loss from LightGCN~\cite{light_gcn} for graph encoder). Additionally, we introduce an auxiliary objective $\mathcal{L}_{f}$ based on fused behavioral representation:
\begin{equation}
    \mathcal{L}_{f} = \sum_{u \in \mathcal{U}}{
    - \log 
        \frac{
        \exp^{(e^u_f)^\top I_e(i^u_p)}
        } {
        \sum_{i_n \in \text{Sample}(u)} { \exp^{ (e^u_f)^\top I_e(i_n) } } + \exp^{(e^u_f)^\top I_e(i^u_p)} 
        }
        },
\end{equation}
where $i^u_p$ is the next interacted item for user $u$, and $i_n$ is a randomly selected negative item ID outside of $S_u$. Also, this representation is used for computing item relevance scores by using the dot-product between this embedding and item embeddings taken from $I_{e}$.

Finally, we introduce contrastive loss $\mathcal{L}_{c}$ between local and global enriched item representations $E^u_l$ and $E^u_g$, aiding the model to make a distinction between direct and indirect dependencies:
\begin{equation}
    \mathcal{L}_c = \sum_{u \in \mathcal{U}}{\sum_{i = 1}^{c}{
    - \log \frac{\exp^{(E^u_l)^\top_{i} (E^u_g)_{i}}}{\sum\limits_{j = 1}^{c}{\exp^{ (E^u_l)^\top_{i} (E^u_g)^\top_{j}}}}
    }},
\end{equation}
where $(E^u_g)_{i}$ and $(E^u_l)_{i}$ -- global and local enriched representations of $i$-th item in the sequence respectively. 

The overall objective is given as follows:
\begin{equation}
    \mathcal{L}(\theta) = 
    \alpha \mathcal{L}_{l} + 
    \beta \mathcal{L}_{g} + 
    \gamma \mathcal{L}_{f} + 
    \delta \mathcal{L}_{c},
    \label{eq:final_loss}
\end{equation} 
where $\alpha,\, \beta,\, \gamma,\, \delta$ are model hyper-parameters determining each objective's influence on the overall model objective.

\section{Experimental setup}

\paragraph{Datasets}

We experiment on four benchmark datasets: \emph{Amazon-Beauty}, \emph{Amazon-Clothing}, \emph{Amazon-Sports}, and \emph{MovieLens-1M}. Data preprocessing follows standard procedure: treating all interactions as positives, filtering out users and items with fewer than \emph{5} interactions, and sorting user sequences by interactions timestamp. Table~\ref{tab:datasets} summarizes statistics after preprocessing.
\begin{table}[h!]
\caption{Statistics of the experimental datasets}
    \begin{tabular}{ c c c c c }
\hline
 \textbf{Dataset} & \textbf{\#Items} & \textbf{\#Users} & \textbf{\#Interactions} & \textbf{Avg. length}\\ 
\hline
Beauty & 12,101 & 22,363 & 198,502 & 8.876 \\ 
Clothing & 23,033 & 39,387 & 278,677 & 7.075 \\ 
Sports & 18,357 & 35,598 & 296,337 & 8.325 \\ 
  ML-1M & 3,706 & 6,040 & 1,000,209 & 165.598 \\ 
 \hline
\end{tabular}
    \label{tab:datasets}
\end{table}

\paragraph{Evaluation setting}

We adopt the leave-one-out strategy as in prior research~\cite{sasrec, bert4rec} for train-validation-test splits: the most recent item is designated as the test target, the second most recent as the validation target (used for best hyper-parameters search), and the remaining items used for training. To avoid unintended biases, we do not perform target item subsampling and rank ground truth items against the entire item catalog. Model performance is assessed using \emph{HR@n} and \emph{NDCG@n} metrics with $n = 5,\, 10$.

\begin{table*}[ht]
\centering
\caption{Experimental results on benchmark datasets}
\begin{tabular}{cl|ccc|cc|c|cc}
\hline
Dataset & Metric & SASRec & BERT4Rec &  SASRec+ & NGCF & LightGCN & MCLSR & MRGSRec & Uplift (\%) \\
\hline
\multirow{4}{*} {Beauty} 
& HR@5 & 0.0352 & 0.0182 & \underline{0.0426} & 0.0130 & 0.0154 & 0.0371 & \textbf{0.0441} & \textit{3.52} \\
& HR@10 & 0.0611 & 0.0412 & \underline{0.0663} & 0.0215 & 0.0253 & 0.0614 & \textbf{0.0681} & \textit{2.71} \\
& NDCG@5 & 0.0214 & 0.0183 & \underline{0.0289} & 0.0070 & 0.0079 & 0.0212 & \textbf{0.0301} & \textit{4.15} \\
& NDCG@10 & 0.0273 & 0.0254 & \underline{0.0343} & 0.0094 & 0.0098 & 0.0275 & \textbf{0.0349} & \textit{1.75} \\
\hline
\multirow{4}{*} {Clothing} 
& HR@5 & 0.0159 & 0.0137 & \underline{0.0207} & 0.0077 & 0.0089 & 0.0168 & \textbf{0.0221} & \textit{6.76} \\
& HR@10 & 0.0271 & 0.0206 & \underline{0.0339} & 0.0133 & 0.0156 & 0.0287 & \textbf{0.0352} & \textit{3.83} \\
& NDCG@5 & 0.0090 & 0.0078 & \underline{0.0114} & 0.0041 & 0.0043 & 0.0093 & \textbf{0.0121} & \textit{6.15} \\
& NDCG@10 & 0.0116 & 0.0105 & \underline{0.0155} & 0.0059 & 0.0071 & 0.0117 & \textbf{0.0162} & \textit{4.52} \\
\hline
\multirow{4}{*} {Sports} 
& HR@5 & 0.0216 & 0.0178 & \underline{0.0275} & 0.0115 & 0.0133 & 0.0233 & \textbf{0.0289} & \textit{5.09} \\
& HR@10 & 0.0324 & 0.0322 & \underline{0.0417} & 0.0155 & 0.0179 & 0.0318 & \textbf{0.0434} & \textit{4.08} \\
& NDCG@5 & 0.0135 & 0.0102 & \underline{0.0176} & 0.0068 & 0.0072 & 0.0134 & \textbf{0.0186} & \textit{5.68} \\
& NDCG@10 & 0.0167 & 0.0153 & \underline{0.0218} & 0.0073 & 0.0080 & 0.0168 & \textbf{0.0227} & \textit{4.12} \\
\hline
\multirow{4}{*} {ML-1M} 
& HR@5 & 0.1067 & 0.0745 & \underline{0.1244} & 0.0439 & 0.0485 & 0.1077 & \textbf{0.1312} & \textit{5.47} \\
& HR@10 & 0.1804 & 0.1333 & \underline{0.2133} & 0.0864 & 0.0921 & 0.1813 & \textbf{0.2248} & \textit{5.39} \\
& NDCG@5 & 0.0639 & 0.0439 & \underline{0.0761} & 0.0286 & 0.0307 & 0.0634 & \textbf{0.0829} & \textit{8.94} \\
& NDCG@10 & 0.0918 & 0.0623 & \underline{0.1065} & 0.0477 & 0.0549 & 0.0921 & \textbf{0.1134} & \textit{6.48} \\
\hline
\end{tabular}
\label{tab:results}
\end{table*}

\paragraph{Baseline Algorithms}

% Since MRGSRec is related to both graph-based and sequential-based approaches, we adopt baseline algorithms from both fields to perform a comprehensive comparison. Specifically, we consider the following baseline models: \textbf{SASRec}~\cite{sasrec}, \textbf{BERT4Rec}~\cite{bert4rec}, and \textbf{SASRec+}~\cite{sasrec_ce} as sequential approaches, \textbf{NGCF}~\cite{ngcf} and \textbf{LightGCN}~\cite{light_gcn} as graph-based models. Moreover, to further assess the efficacy of our model, we compare it with an approach that integrates both data representations, namely \textbf{MCLSR}~\cite{mclsr}. For the implementation of MRGSRec we used sequential encoder and loss function from \textbf{SASRec+} and graph encoder with downstream losses from \textbf{LightGCN} since we wanted to see whether MRGSRec would be able to boost the performance of the best representatives from each family.

We compare MRGSRec with baselines from both sequence-based and graph-based approaches. These models include \textbf{SASRec}, \textbf{BERT4Rec}, and \textbf{SASRec+} for sequence-based, and \textbf{NGCF} with \textbf{LightGCN} for graph-based models. Additionally, we evaluate MRGSRec against \textbf{MCLSR}, an approach integrating both data representations. During this comparison, MRGSRec implementation incorporates the sequential encoder and loss function from \textbf{SASRec+}, and the graph encoder with objective function from \textbf{LightGCN} to explore its boost upon these strong baselines.

\section{Results and Discussion}

Table~\ref{tab:results} reports the overall performance compared with baselines. The obtained results highlight a more pronounced relative uplift for $n = 5$ compared to $n = 10$. Although there is also an uplift for $n = 20$ (not reported here for brevity), it is observed to be of a lesser extent. We conclude that the model excels at promoting positive items to the top of recommendation lists, but revealing additional positive examples undetected by other models remains challenging.

Our analysis shows that the ML-1M dataset, with a higher density of interactions, demonstrates the most significant relative improvement in the NDCG metric. This suggests that our model may enhance performance when dealing with more interactions, possibly due to graph models performing better for frequently consumed items. This observation could explain the comparatively lower uplift metric values observed for the Beauty dataset.

In this paper, we have presented a novel end-to-end framework for leveraging sequence-based and graph-based data representations into a single model to improve the representation of user behaviors. The proposed framework, MRGSRec, constructs two independent representations from sequence-based and graph-based encoders that can capture different patterns in the input data. Our experiments on classical benchmarks demonstrate the effectiveness of our approach against diverse families of state-of-the-art methods.

\bibliographystyle{ACM-Reference-Format}
\bibliography{main}

\end{document}